\documentclass{ieeeaccess}
\usepackage[english]{babel}
\usepackage{cite}
\usepackage{amsmath,amssymb,amsfonts}
\usepackage{graphicx}
\usepackage{textcomp}
\usepackage{listings}
\usepackage{enumitem}
\usepackage{multirow}
\usepackage{microtype}
\usepackage{pifont}
\usepackage{makecell}
\usepackage{hyperref}
\usepackage{xurl}

\def\BibTeX{{\rm B\kern-.05em{\sc i\kern-.025em b}\kern-.08em
    T\kern-.1667em\lower.7ex\hbox{E}\kern-.125emX}}

\lstdefinelanguage{Tamarin}
{
  morekeywords={rule, lemma, In, Out, let, constrain, functions, query, fact, and, or, not, K, Fr, Is_Agent, New_Client_TLS_Session, New_Server_TLS_Session, Client_Starts_TLS, SP_Initialized, Is_Resource, Is_Registered, Ltk, New_Id, Start_SP, SP_Resource_Request, Https_Response, SP_State_1, All, Ex, Client_Params, Compromise_Agent, Register_SP_at_IdP, Register_Resource_at_SP},
  sensitive=true,
  morecomment=[l]{//},
  morecomment=[s]{/*}{*/},
  morestring=[b]',
}

\newcommand{\yes}{\textcolor{green}{\checkmark}}
\newcommand{\no}{\textcolor{red}{\ding{55}}}

\newcommand{\samlmsg}[1]{\texttt{<#1>}}
\newcommand{\msg}[1]{\texttt{#1}}
\DeclareRobustCommand{\xjoinrel}{\mathrel{\mkern-3.5mu}}
\newcommand{\ltr}[3]{\mathrel{[#1]}\xjoinrel\mathrel{-}\xjoinrel\mathrel{[#2]}\xjoinrel\rightarrow\xjoinrel\mathrel{[#3]}}

\begin{document}
\history{Date of publication xxxx 00, 0000, date of current version xxxx 00, 0000.}
\doi{10.1109/ACCESS.2023.0322000}

\title{Towards automated formal security analysis of SAML V2.0 Web Browser SSO standard --- the POST/Artifact use case}
\author{\uppercase{Zvonimir Hartl}\authorrefmark{1} and
\uppercase{Ante Derek\authorrefmark{2} \IEEEmembership{Member, IEEE}}}
\tfootnote{This research has been supported by the European Regional Development Fund under grant agreement PK.1.1.10.0007 (DATACROSS)}

\address[1]{University of Zagreb, Faculty of Electrical Engineering and Computing, 10000 Zagreb, Croatia (e-mail: zvonimir.hartl@fer.hr)}
\address[2]{University of Zagreb, Faculty of Electrical Engineering and Computing, 10000 Zagreb, Croatia (e-mail: ante.derek@fer.hr)}

\markboth
{Author \headeretal: Preparation of Papers for IEEE TRANSACTIONS and JOURNALS}
{Author \headeretal: Preparation of Papers for IEEE TRANSACTIONS and JOURNALS}

\corresp{Corresponding author: Zvonimir Hartl (zvonimir.hartl@fer.hr).}

\begin{abstract}
Single Sign-On (SSO) protocols streamline user authentication with a unified login for multiple online services, improving usability and security.
One of the most common SSO protocol frameworks --- the Security Assertion Markup Language V2.0 (SAML) Web SSO Profile --- has been in use for more than two decades, primarily in government, education and enterprise environments.
Despite its mission-critical nature, only certain deployments and configurations of the Web SSO Profile have been formally analyzed.
This paper attempts to bridge this gap by performing a comprehensive formal security analysis of the SAML V2.0 \emph{SP-initiated SSO with POST/Artifact Bindings} use case. 
Rather than focusing on a specific deployment and configuration, we closely follow the specification with the goal of capturing many different deployments allowed by the standard.
Modeling and analysis are performed using Tamarin prover --- state-of-the-art tool for automated verification of security protocols in the symbolic model of cryptography. 
Technically, we build a meta-model of the use case that we instantiate to eight different protocol variants.
Using the Tamarin prover, we formally verify a number of critical security properties for those protocol variants, while identifying certain drawbacks and potential vulnerabilities.
\end{abstract}


\titlepgskip=-21pt

\maketitle

\section{Introduction}
\label{sec:introduction}
\PARstart{I}{ncreasing} number of online services are requiring user authentication, with passwords still being the prevalent authentication method. 
Accessibility issues related to unique strong passwords naturally lead to password reuse, which, in turn, makes services vulnerable to various attacks~\cite{credential-stuffing,credential-stuffing2}, degrading the overall security on the Internet.  
Simultaneously, the burden on service providers to deploy, maintain and safeguard the authentication infrastructure further compounds the complexity.
In response to these challenges, the concept of Web Single Sign-On (SSO) emerged at the start of this millennium.
SSO protocols enable exchange of authentication and authorization data between trusted identity providers (IdPs) and service providers (SPs), enabling users to authenticate once at an IdP and access services across multiple SPs.

The Security Assertion Markup Language (SAML) version 2.0 was adopted as a standard in 2005 by the Organization for the Advancement of Structured Information Standards (OASIS), replacing version 1.0 adopted in 2002.
Since that time, SAML V2.0 has emerged as the predominant SSO framework in government, education and enterprise environments, while its competitors such as OAuth 2.0 \cite{oauth_rfc} and OpenID Connect (OIDC) \cite{OIDC_spec_doc} are more commonly used on the open Web.

SSO frameworks ease the burden of authentication on SPs, but also impose stricter security requirements for IdPs and the SSO framework itself, as they are a single point of failure --- a compromised IdP or a weakness in the SSO system can lead to compromises of sensitive user data across different SPs.
For example, in the \emph{Sunburst} supply chain attacks \cite{fireeye_sunburst, microsoft_sunburst} , after the initial compromise, the attackers obtained the trusted signing certificate used for signing SAML assertions, and used it to access all services within the targeted enterprise.

There have been numerous efforts in empirical security analysis of OAuth 2.0 \cite{Yang2016_OAuth, wang_oauth2013, sunOauth2012, OAuth_mobile, OAuth_mobile2, OAuth_mobile3}, OIDC  \cite{mladenov2016securitySSO_oidc} and SAML  \cite{Mainka2017_func_sec_OIDC_36, on_break_saml, Mainka2014_SAML_saas,eidas_sec}, resulting in  improved protocol implementations and specifications.
Whilst the empirical security analysis can identify flaws and improve security, it can not guarantee the absence of weaknesses.
On the other hand, formal security analysis can offer valid proofs, sometimes machine verified or even machine generated, that the mathematical model of the protocol holds desired security properties. 
Of course, the security guarantees provided by a deployed system depend on the faithfulness of the mathematical model to the specification and implementation.

Formal analysis of security protocols usually employs either the \emph{symbolic} or \emph{computational} approach. Symbolic analysis treats cryptographic primitives as abstract operations and focuses on the logical flow of information within the protocol.
On the other hand, computational model more closely resembles the real world --- the messages are bitstrings, the adversary is capable of arbitrary computation (while using reasonable resources), and cryptographic primitives come with precise complexity-theoretic security guarantees.
Symbolic analysis is more automation-friendly and hence more suitable for complex protocols. However, protocol proved secure in the symbolic model might still be vulnerable to real world attacks. While there are tools for automated analysis in the computational model \cite{easy_crypt, crypto_verif}, the analysis is significantly more complex and still infeasible for larger protocols. 

This research uses the Tamarin prover~\cite{Tamarin_paper}, a state-of-the-art tool for automated symbolic protocol analysis and --- alongside ProVerif~\cite{proverif} --- one of the most prominent tools in the field.
Tamarin is well suited for modelling and analysing intricate security protocols due to its support for complex protocol flows, branching, loops, and mutable global state.
These features are essential for modeling SAML V2.0 participants and their interactions, as well as the trust establishment steps.
Finally, the choice of Tamarin was also influenced by the authors' prior experience with the tool; however, we see no fundamental obstacles preventing other tools, such as ProVerif~\cite{proverif}, from being used to comprehensively model and analyze SAML V2.0 as well.

Our research aims to conduct a systematic symbolic formal analysis and verification of the security properties associated with the SAML V2.0 SSO Profile.
Hence, we would ideally formally verify that SAML SSO implementations adhering to the technical specifications outlined within the official SAML V2.0 standard documentation \cite{saml_profiles, saml_sec_privacy} do achieve the intended security properties: Secrecy, Authentication, Authenticity and Freshness. Of course, whenever the security properties are not achieved, our goal is to identify and evaluate attacks that break the said properties, and to suggest mitigations and protocol improvements.

Despite its mission-critical nature and wide adoption, SAML V2.0 has received very little attention from the formal methods' community, with the work by Armando et al.~\cite{saml_google_2008,SAML_Armando_2013,SAML_Armando_2011} being the only attempt to formally analyze systems based on the SAML framework.
Their research, however, is exclusively focused on the so-called \emph{Redirect/POST use case} which is only one of several patterns for building Web SSO with SAML (albeit the most common one in use today).
Furthermore, Armando et al. only considered one specific deployment configuration used by Google, while SAML standard allows for multiple protocol variants and features.
There are no attempts at formal analysis of other use cases, let alone to model the entire specification.

One of the obstacles for formal analysis is that SAML is not itself a security protocol --- it is a set of building blocks that can be combined, configured and implemented in many different ways to build the protocol that achieves the desired SSO functionality.
Even when we fix the specific SSO use case (e.g., the  \emph{Redirect/POST use case} in this paper), there are at least two additional sources of flexibility  --- \emph{optional features} and \emph{under-specified mechanisms}. 
While such flexibility is sometimes purely functional, there are cases where both optional features and under specified mechanisms do have an influence on the resulting security properties.

While the SP-initiated Redirect/POST binding is more common, in this paper we focus on the POST/Artifact use case.
This is the first analysis of an important and widely used SAML Web SSO SP-initiated POST/Artifact use case (three EU countries mandate support for this use case as part of their eID services \cite{eID_overview}).
Moreover, SAML V2.0 documentation positions POST/Artifact use case as an alternative to Redirect/POST use case \cite{saml_tech}.
Despite its lower adoption rate due to implementation complexity on both the SP and IdP sides, the assumption is that POST/Artifact should offer potential security benefits since the authentication response is not directly transmitted within the user's browser redirect.

In this paper, we are not trying to model a specific implementation --- we attempt to model the \emph{specification}, accounting for the flexibility allowed by the standard.
We develop a straightforward methodology that, first, identifies \emph{sensitive optional features} and \emph{sensitive under-specified mechanisms}.
Such features and mechanisms can be instantiated in several ways, with each instantiation resulting in what we call a \emph{protocol variant}.
An example of optional features are digital signatures of authentication requests, which are recommended (but not required) by the SAML standard.
Furthermore, we perform automated analysis of protocol variants, yielding security proofs or falsifications of desired properties.
Finally, the result of the formal analysis validate the methodology and demonstrate that protocol variants do indeed have different security features.

There are multiple benefits to this approach.
Firstly, we can use automated formal analysis tools to examine the exact impact of each feature on the overall security guarantees provided by the protocol, as SAML standards only gives vague guidelines.
Using the example of optional signatures, the standard~\cite{saml_core} says \emph{``The SAML request
MAY be signed, which provides both authentication of the requester and message integrity.''} --- it is unclear what exactly (if anything) can go wrong in systems configured not to require signed authentication requests.
Our analysis gives an answer to this question --- for the POST/Artifact use case, adding signatures to authentication requests does not have a significant influence on the security of the use case in the symbolic model.

The analysis presented in this paper, while significantly more comprehensive than the related work~\cite{saml_google_2008,SAML_Armando_2013,SAML_Armando_2011}, still falls short of fully modelling the use case --- we only consider three degrees of freedom, and model only two options for each. Even with automated formal analysis, we are still unable to analyze \emph{all possible} protocol variants due to the complexity explosion.

In order to perform the analysis, we needed to overcome a number of technical challenges related both to SAML itself and modeling with the Tamarin prover. 
For example, we needed to examine various SAML implementations to identify how exactly are under-specified mechanisms implemented in practice. Also, we carefully model trust establishment steps to account for arbitrary complex system and trust relationships. Finally, we  develop a novel method of modelling TLS communication with Tamarin prover since we found existing approaches unsuitable for this purpose.

\paragraph*{Contributions}
In summary, the contributions of this paper are as follows:
\begin{itemize}
    \item We build the first formal model of the SAML 2.0  \emph{POST/Artifact} use case. The model is based on the SAML standard rather than a particular deployment, and created using the methodology developed in this paper. The result is a meta-model of the POST/Artifact use case, that we instantiate to eight different protocol variants.
    \item While the developed methodology is simple and straightforward, we consider it an independent contribution and an important first step towards comprehensively modeling and analyzing SAML Web SSO profile in its entirety.
    \item We give automated formal proofs of the most important security properties of the eight protocol variants, including multiple authentication, secrecy and freshness properties. The properties hold in a general setting with an unbounded number of protocol participants and sessions, and a Dolev-Yao~\cite{dolev_yao} adversary that can compromise arbitrary participants.
    \item We identify a weakness in all protocol variants --- the \emph{artifact-session confusion attack}. While the discovered attack does not seem to have an impact on the security of deployed systems, it highlights the necessity of closely following the SAML assertion processing rules. We suggest a simple fix, and formally verify that it does indeed mitigate the discovered attack. 
    \item We include the Tamarin code and the scripts~\cite{github_code_ref} needed to reproduce the results as a submission artifact, and plan to make the complete model publicly available upon paper publication.
\end{itemize}

\paragraph*{Organization of the paper}
First, we discuss related work in Section~\ref{section:related}.
A brief description of the SAML standard and the SAML Web SSO Profile is given in Section~\ref{section:background}. Then, we present the methodology and the formal model of the POST/Artifact use case in Section~\ref{section:formal_model}.
In Section~\ref{section:properties}, we present the security properties and the result of their verification, with the discussion summarizing the results in Section~\ref{section:discussion}. 
Finally, we conclude in Section~\ref{section:conclusion}.

\section{Related work}
\label{section:related}
Considering that SAML-based SSO systems are widely used, especially in government, corporate and educational settings, there are surprisingly few efforts to formally analyze these systems.
Our research is the first attempt at a formal analysis of the POST/Artifact bindings use case, as well as the first attempt to analyze the technical specification rather than a particular deployment.  

The most notable example of a formal analysis of a SAML SSO use case is the research of Armando \emph{et al.}~\cite{saml_google_2008,SAML_Armando_2013,SAML_Armando_2011}.
They model and analyze the Redirect/POST use case, as well as an actual deployment by Google, using a SAT-based model checker SATMC~\cite{satmc1st}.
They consider two security properties --- authentication based on non-injective agreement and the resource secrecy.
The most significant contribution is the discovery of a crucial flaw in the Google's deployment --- the SP identifier was left out from the \samlmsg{AuthnRequest} message, leading to an attack breaking both security properties considered.
The authors also note that the tool fails to find similar attacks when the SP identifier is included in the message, as required by the SAML specifications. 

In the only other relevant attempt at a formal analysis of SAML use cases~\cite{formalising_all_SSO}, Ferdous and Poet analyze the same use case as in~\cite{SAML_Armando_2011}, and create a simpler model, but one that includes an explicit user authentication step and a rudimentary \emph{digital identity} model.
The security properties are again similar to~\cite{SAML_Armando_2011} and include basic authentication and secrecy properties.
They use the High-Level Protocol Specification Language (HLPSL) to specify the protocols and the Automated Validation of Internet Security Protocols and Applications (AVISPA) tool with the On-the-fly Model-Checker (OFMC) backend for analysis. 
Besides the SAML use case, they also investigate other SSO protocols, namely OpenID Connect and OAuth.
The analysis by the tool discovered no attacks, but the authors do not make explicit claims on the exact setting in which the properties hold.

\begin{figure*}[t]
    \centering
    \includegraphics{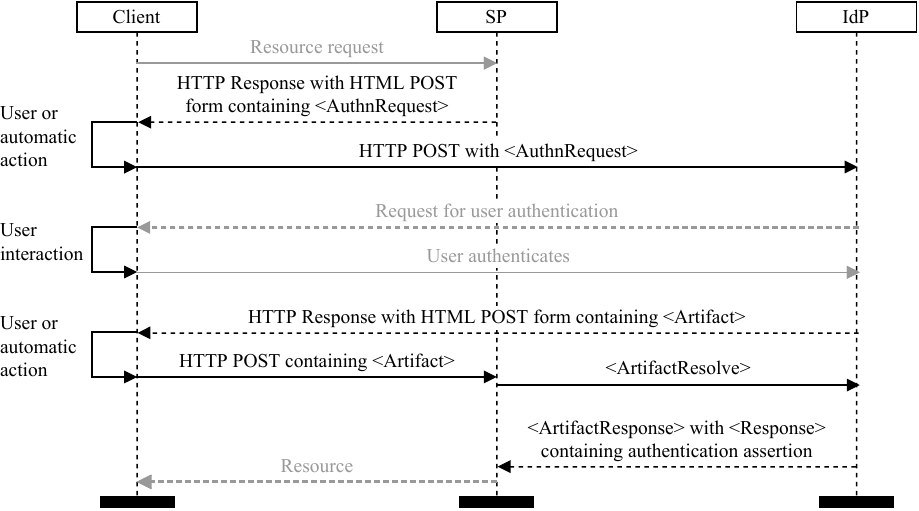}
    \caption{An example protocol execution of SAML V2.0 SSO SP-Initiated SSO with POST/Artifact bindings use case. Steps outside the scope of SAML Web Browser SSO Profile are grayed out.}
        \label{SAML_flow}
\end{figure*}

Our work differs from the two efforts mentioned above in several aspects. 
First, we focus on a different SAML use case. 
More importantly, we build a much more detailed and comprehensive model of the system (the protocol setup, the protocol variants, a more general adversary that can compromise any participant,  etc.) and consider a larger number of relevant security properties.
Finally, the bounded model checker used in~\cite{SAML_Armando_2011} only guarantees the absence of attacks assuming a small fixed number of participants and sessions.
In contrast, our positive results hold in a much more general setting with an unbounded number of protocol participants and sessions.

In the more general area of formal analysis of Web SSO systems and standards, Fett \emph{et al.} develop an expressive Dolev-Yao style model of the web infrastructure (the \emph{FKS model}) and use it to perform a comprehensive formal analysis of OAuth~\cite{FKS_OAuth} and OpenID Connect~\cite{FKS_OIDC}. The analysis is not machine-verified, but others have replicated their models and confirmed the results using the Tamarin prover \cite{tamarin_oidc_dunki_masters, tamarin_oidc_xenia_masters}.

Finally, there are several important efforts to analyze particular implementations and deployments of SAML-based systems and detect low-level attacks. Somorovsky \emph{et al.} analyze 14
major SAML frameworks and show that 11 of them have vulnerabilities related to XML signature handling~\cite{on_break_saml}.
Mainka \emph{et al.} perform security analysis of 22 SaaS cloud providers that use a SAML-based SSO~\cite{Mainka2014_SAML_saas}.
They identify  eight different SAML-related attack vectors and show that 90\% of the providers are vulnerable to at least one attack.
Finally, Engelbertz \emph{et al.} identify numerous vulnerabilities in SAML-based eID implementations in European Union countries~\cite{eidas_sec}.
They also develop the EsPReSSO tool for automatic security analysis of SAML-based SSO systems.

\section{Background}
\label{section:background}

\subsection{SAML Framework}
\label{section:saml}
The Security Assertion Markup Language version 2.0 (in the rest of paper, the abbreviation SAML will denote SAML version 2.0) is a technical framework for exchanging security information between partners~\cite{saml_tech}.
It is an open standard developed by the Security Services Technical Committee under OASIS.
The brief technical overview of the SAML standard is given in the official document~\cite{saml_tech}, whereas the standard itself is defined in a series of technical documents~\cite{saml_core, saml_bindings, saml_profiles}.

SAML framework defines the following basic building blocks: assertions, protocols, bindings and profiles.
\emph{Assertions} define the syntax, semantics and processing rules for security statements.
\emph{Protocols} define syntax, semantics and high-level processing rules for requesting and delivering assertions.
Specification for both assertions and protocols includes various application-level cryptographic mechanisms (e.g., digital signatures for ensuring authenticity of authentication requests, public-key encryption for protecting the confidentiality of assertion data).
\emph{Bindings} map the abstract protocols to specific delivery mechanisms (e.g., SOAP, HTTP Redirect, HTTP POST, artifact).
The bindings' specification also mandates a transport layer security mechanism --- the TLS protocol --- in certain situations.
Finally, \emph{profiles} combine SAML assertions, protocols and bindings with an additional set of rules in a system that aims to provide a specific functionality (e.g., a SSO mechanism on the web).

Furthermore, multiple SAML profiles can be configured and combined in different ways to implement specific identity federation \emph{SAML use cases}~\cite{saml_tech}.
For example, one common use case is \emph{SP-Initiated SSO with Redirect/POST Bindings} (the ``Redirect/POST'' use case, in short). Finally, each \emph{SAML deployment} for a specific SAML use case requires further configuration on all levels (assertions, protocols, bindings and profiles), trust establishment between parties and implementation of system parts that are left unspecified by the SAML framework (e.g., the authentication of end users).

\begin{table*}[t]
\centering
\renewcommand{\arraystretch}{1.2}
\begin{tabular}{p{1.79cm}p{7.38cm}p{7.28cm}}
\textbf{Message} & \textbf{Relevant contents} & \textbf{Protocol step description} \\ 
\hline 
Resource request & Resource URL & User attempts to access resource on SP, no valid session exists. SP stores the requested resource URL using the RelayState mechanism. \\

Authentication Request & IdP, RelayState, AuthnRequest(OptionalSignature, id\_authnrequest, SP, IdP) & SP sends user an HTML form containing encoded SAML AuthnRequest (SAMLRequest) and RelayState. User submits form or script auto-submits to IdP's SSO service.\\

Request for user authentication &  Authentication challenge & IdP checks for existing session or prompts user for credentials. \\

User authenticates & User credentials & IdP verifies user credentials (if needed) and establishes session. \\

Artifact & RelayState, Artifact(messageHandle) & IdP creates a SAML Response containing user assertion. IdP creates an artifact referencing the Response and redirects user's browser with the artifact. \\

Artifact Resolve & ArtifactResolve(id\_artifactresolve, SP, messageHandle, ArtifactResolveSignature) & SP sends the received Artifact to the IdP's Artifact Resolution Service to retrieve the original SAML Response.\\

Response & \makecell[tl]{ArtifactResponse( IdP, id\_artifactresolve,
\\ \quad Response( ResponseSignature, id\_response, IdP, id\_arthrequest, \\ \qquad Assertion( id\_assertion, IdP, Audience(SP), assertion\_secret, \\ \qquad\quad Subject(clientID, id\_authnrequest) ) ) )} & \makecell[tl]{The IdP sends to the SP the SAML Response referenced by\\ received artifact.} \\

Resource & Resource & The SP extracts the SAML Response from the ArtifactResponse validates it (if not already signed), and processes the embedded assertion. Based on the assertion, the SP creates a local security context for the user. Finally, the SP retrieves the originally requested resource URL from the RelayState and redirects the user's browser to access that resource.  \\
\hline

\end{tabular}
\caption{Description of SAML POST/Artifact messages from a typical execution flow.}
\label{table:message_desc}
\end{table*}

\subsection{Web Browser SSO Profile}
\label{section:sso}
In the SAML \emph{Web Browser SSO Profile}, an \emph{end user} uses a web browser client to access a resource at a \emph{service provider} (SP) while verifying its identity to an \emph{identity provider} (IdP).
On a high level of abstraction, when the user attempts to access a resource, the SP hosting the resource generates and communicates an \emph{authentication request} to an IdP.
The IdP, after verifying the end user identity, communicates a \emph{response} back to SP containing an \emph{assertion} describing the identity of the end user as well as other metadata.
This profile assumes that SPs and IdPs have established trust and that IdPs have methods of authenticating end users.
However, the details of trust establishment and user management/authentication are outside the scope of the SAML standard.

The communication of authentication requests and responses depends on the protocol bindings used.
For example, in the Redirect/POST use case, an HTTP Redirect (over TLS) is used to deliver the authentication request from the SP to the IdP via the web browser client, whereas the HTTP POST mechanism is used to deliver back the response, again via the web browser client. 

With \emph{HTTP artifact binding} the requests and/or responses can be communicated by reference using so-called \emph{artifacts}. In that case, another SAML profile --- the \emph{Artifact Resolution Profile} --- is additionally used to resolve the artifact into an actual  request and/or response. 

\subsection{SP-initiated SSO with POST/Artifact Bindings}
\label{section:post_artifact}
In this paper, we model and analyze the \emph{SP-initiated SSO with POST/Artifact Bindings} use case as defined in ~\cite{saml_profiles, saml_tech}.
In this use case, the authentication requests are sent using the HTTP POST binding, whereas the responses from IdPs are sent back using the HTTP artifact binding.
Furthermore, the flows are initiated by an end user using a web browser client to request a resource at SP (SAML also supports IdP initiated SSO).
Note that, from the symbolic protocol analysis perspective, HTTP Redirect and POST delivery mechanisms are equivalent. Still, we will refer to the use case as ``POST/Artifact'' in the rest of the paper to be consistent with the SAML conventions.

A high level execution flow is given in Fig.~\ref{SAML_flow}, whilst each message contents are given in Table~\ref{table:message_desc}.
End user through the web browser client attempts to access a resource at SP.
SP responds with a script that initiates an HTTP POST request to the IdP containing the \samlmsg{AuthnRequest} structure --- a SAML message consisting of, amongst other data, a unique id, SP identifier and IdP identifier. 
This \samlmsg{AuthnRequest} can be accompanied by a \msg{RelayState} message which holds information about end user's initial request (this is more discussed in Section~\ref{section:methodology}).
IdP receives the \samlmsg{AuthnRequest} and tries to authenticate the end user.
After successful authentication, IdP builds the SAML \samlmsg{Response} message, but does not deliver it via the web browser client.
Instead, it creates an \samlmsg{Artifact} SAML message containing an unique ID associated with the specific \samlmsg{Response} message.
This \samlmsg{Artifact} is delivered to the SP via the web browser using either HTTP Redirect or POST mechanism.
After receiving the \samlmsg{Artifact}, SP builds the \samlmsg{ArtifactResolve} message and sends it directly using SOAP binding to IdP.
After successfully verifying the \samlmsg{ArtifactResolve} request, IdP responds  with a \samlmsg{Response} containing an assertion.
Finally, the SP verifies the assertion and provides the requested resource to the end user.

\subsection{Tamarin prover}
\label{section:tamarin}
The Tamarin prover~\cite{Tamarin_paper, BSchmidt_thesis, SMeier_thesis} is a state-of-the-art tool for symbolic modeling and automated analysis of security protocols. It was successfully used to analyze, break, and improve many real-world security protocols such as TLS 1.3~\cite{tamarin_tls}, authentication protocols for 5G networks~\cite{tamarin_5gA}, and WPA2~\cite{tamarin_wpa}.

In Tamarin, protocols are modeled by multiset-rewriting \emph{rules}, each representing an atomic action.
A rule is of the form $id{:}\ltr{l}{a}{r}$, where $l,a,r$ are sets of \emph{facts} --- \emph{pre-conditions}, \textit{action} facts and \emph{post-conditions}, respectively.
In a typical rule, pre-conditions will correspond to a message an agent receives and its internal state, whereas the post-conditions will correspond to the messages the agent sends and the changes to its internal state.

Security properties are most often specified using \emph{guarded first-order logic formulas} involving actions facts, that are interpreted over execution traces. Tamarin also supports equivalence properties, but we use trace properties exclusively in this paper. The verification procedure tries to automatically verify or falsify the formula in a \emph{constraint system} containing user-defined rules, built-in adversary rules and model \emph{restrictions}. 
The automated procedure can be further refined by the user by supplying a proof-guiding \emph{oracle} --- a sort of heuristics for optimizing automated proving.
The property, if verified, holds in a general setting with an unbounded number of protocol participants and sessions unless, of course, different assumptions are explicitly built into the model or the properties.

\section{Formal model} 
\label{section:formal_model}
Now, we present the formal model of the POST/Artifact use case as defined in ~\cite{saml_profiles, saml_tech}.
To obtain the formal model of the \emph{system}, we need to model the behaviour of the participants defined by the standards --- namely the SPs and IdPs, but also the end users participating in the protocol by using a web browser client.
Furthermore, we need to explicitly model the trust establishment steps between protocol participants that are implicit in the SSO use case.
Since the use case assumes TLS connections, which are not provided by default in the Tamarin prover, we build primitives that model web request-response communication over one-way authenticated TLS.
Finally, we need to model the adversary that, besides the usual Dolev-Yao capabilities, can compromise other protocol participants (SPs, IdPs, and clients) and communicate over TLS with honest participants.

\subsection{Methodology and Model Variants}
\label{section:methodology}
\begin{table*}
\begin{center}
\renewcommand{\arraystretch}{1.1}
\begin{tabular}{p{4cm}>{\centering\arraybackslash}m{1.1cm}>{\centering\arraybackslash}m{1.1cm}p{9cm}}
\textbf{SAML element full path} &
\textbf{Required} &
\textbf{Security sensitive} &
\textbf{Element description} \\
\hline
\makecell[l]{<AuthnRequest>\\ \quad <Subject>} & \no & \yes  & Specifies the subject of the authentication request or assertion, which identifies the principal that is the subject of the request or assertion. \\
\makecell[l]{<AuthnRequest>\\ \quad<Version>} & \yes & \no  & Specifies the version of the SAML protocol. \\
\makecell[l]{<AuthnRequest>\\ \quad<IssueInstant>} & \yes & \no  & Specifies time at which the request or assertion was issued. \\
\makecell[l]{<AuthnRequest>\\ \quad<ForceAuthn>} & \no & \yes  & Indicates whether the identity provider must authenticate the principal even if the principal has an existing session with the identity provider. \\
\makecell[l]{<AuthnRequest>\\ \quad<IsPassive>} & \no & \yes  & Indicates whether the identity provider must not interact with the principal, such as by prompting the principal for credentials, during the authentication process.\\
\makecell[l]{<Response>\\ \quad<EncryptedAssertion>} & \no & \yes  & Represents an assertion that has been encrypted to protect its confidentiality. \\
\makecell[l]{<Response>\\ \quad<Assertion>\\ \qquad<AuthnStatement>} & \yes & \yes  & Contains information about the authentication of a principal, including the time of authentication and the method of authentication used.\\
\makecell[l]{<Response>\\ \quad<Assertion>\\ \qquad<Subject>\\\quad\qquad<SubjectConfirmationData>\\ \qquad\qquad<OneTimeUse>} & \no & \yes  & Indicates whether the authentication request or assertion can only be used once, to prevent replay attacks.\\
\makecell[l]{<Response>\\ \quad<Assertion>\\ \qquad<Subject>\\ \quad\qquad<SubjectConfirmationData>\\ \qquad\qquad<NotOnOrAfter>} & \yes & \yes  & Specifies the time after which the authentication request or assertion is no longer valid. \\
\hline
\multicolumn{4}{l}{\textbf{Legend:}} \\
\multicolumn{4}{l}{\quad Second column indicates whether the element is required by SAML standard or not. }\\
\multicolumn{4}{l}{\quad Third column indicates whether the element has an impact to the security properties by our judgment. }\\
\end{tabular} 
\caption{Elements that are not implemented in our model which are required by SAML standard or/and relevant to the security properties of protocol.}
\label{table:elements}
\end{center}
\end{table*}

As mentioned before, there are several degrees of freedom in the SAML specifications, and here we describe the methodology by which we converted the specifications to the \emph{core} formal model of SAML and well as the eight \emph{protocol variants} we consider in our analysis.

First, as expected from the complex standard aiming to capture different use cases, there are many optional features. 
These are reflected as optional fields in SAML assertions, or optional steps in SAML protocols.
For example, the IdP \emph{may} include the optional \samlmsg{Conditions} field in the issued assertion. 
Note, however, if such a field is present, the assertion consumer \emph{must} validate the condition before accepting the assertion.

Our approach for optional features was as follows:
\begin{itemize}
\item All required fields and processing rules (i.e., those described with \emph{MUST} key word in the standard) are included in the core model, with the following exception: If the lack of processing rules gives strong evidence that the field plays a purely functional rather than security role, we leave it out of the model. \item No optional fields and processing rules (i.e., those described with \emph{SHOULD} or \emph{MAY} keywords in the standard) are included in the core model. However, if the specification processing rules, or our judgment, indicate that the optional fields could potentially have an impact on the resulting security properties, we include the optional field in the list of \emph{sensitive optional features}. 
\end{itemize}

Example required fields not included in the model are the \samlmsg{Version} and the \samlmsg{Timestamp} of the \samlmsg{Assertion}.
An example sensitive optional feature is the use of \samlmsg{EncryptedAssertion} instead of \samlmsg{Assertion} in authentication responses. This feature is clearly security relevant, and we recognize as such. Note, however, we leave the implementation for future work rather than adding it to the core model right now.  
Table~\ref{table:elements} contains the exhaustive list of all fields of \samlmsg{AuthnRequest} and \samlmsg{Response} \emph{not} included in our core model, with the fields we consider security sensitive. 

The other source of difficulty when converting the standard to the formal model is the lack of specification of various mechanism. 
Here, we tried to assess if the under-specified mechanism plays a purely functional role or if it could potentially have an impact on the security properties of the overall system.
In the latter case, we include the mechanism in the list of \emph{sensitive under-specified mechanisms}.
For example, the SAML standard specifies that ``dereferencing of the artifact using the Artifact Resolution profile \emph{MUST} be mutually authenticated, integrity protected, and confidential``~\cite{saml_profiles}, but does not explicitly specify the mechanisms to achieve these properties.

When deciding how exactly to model an under-specified mechanism, we considered a straightforward canonical implementation and, additionally, examined various SAML implementations to identify how are those under-specified mechanisms implemented in practice.

After identifying the sensitive optional features and sensitive under-specified mechanisms, we decided to pick three of these  and model each in two different ways, yielding eight different \emph{protocol variants}.

Even with automated formal analysis, it was challenging to model and verify \emph{all possible} protocol variants (specific challenges are discussed in Section~\ref{section:verification}). In this paper, we opted to focus on three specific degrees of freedom, described below. Since one of our goals was to investigate if, and how, small changes in the interpretation of the standard impact the security of the complete system, we picked the features and implementations that seemed most likely to have a significant impact, but are also illustrative of the kinds of choices and potential pitfalls in building a standards-compliant SAML deployment.
We acknowledge that, for a truly comprehensive analysis, it is necessary to consider all properties from Table~\ref{table:elements}, and even elements beyond the scope of SAML specification such as the user behavior.
However, such extensions are reserved for future work. 

Technically, protocol variants are implemented by building a \emph{meta model} with simple preprocessor directives enabling us to automatically generate eight models for all combinations of selected features.

\paragraph*{Authentication request signing}
As mentioned before, signatures in \samlmsg{AuthnRequest} messages are optional in the SAML standard.
Informally, signatures guarantee the integrity and authenticity of the \samlmsg{AuthnRequest} messages and \emph{may} be included in deployments where the authentication of the request issuer is required. 
Hence, we created the ``signed'' and ``unsigned'' model variants with signed and unsigned \samlmsg{AuthnRequest} messages in order to precisely determine the impact of those signatures.

\paragraph*{RelayState mechanism}
Some SAML bindings allow the use of the \emph{RelayState} mechanism to preserve state information. 
In a POST/Artifact use case, the SP \emph{may} include the arbitrary \msg{RelayState} data alongside the \samlmsg{AuthnRequest} --- the IdP \emph{must} return the same data along with the protocol response.
Typically, \msg{RelayState} data enables the SP to ``look up'' or reconstruct the original request and take the appropriate action upon receiving the assertion.
We consider the RelayState mechanism a sensitive under-specified mechanism since it could potentially impact the resource authenticity property --- informally, the resource the client is delivered has to be the same resource it originally requested.

We model two different behaviors corresponding to the two different options for the RelayState mechanism in the popular open source SAML framework Shibboleth~\cite{shibboleth_doc}:
\begin{itemize}
    \item In the ''norelaystate'' model variant, no state is stored in the SP and the URL requested by the client is sent as \msg{RelayState} data. When the SP receives the response (the \samlmsg{Artifact} message and the \msg{RelayState} message containing the URL), it assumes that the URL is the original URL requested by the client. This behavior is consistent with the default behavior of Shibboleth Service Provider v3.
    \item In the ''relaystate'' model variants, the SP generates a new nonce, stores it in the session data along with the original URL, and inserts the nonce into the \msg{RelayState} data. This nonce is used to join the original SP session when the response is received. This behavior is equivalent to the ''ss:mem'' value for the Shibboleth Service Provider v3 ''relayState'' option (which is also the recommended option).
\end{itemize}

\paragraph*{SAML identifiers}
SAML specification mandates the use \emph{identifiers} for many SAML messages.
The specification requires that ``\emph{any party that assigns an identifier MUST ensure that there is negligible probability any other party will accidentally assign the same identifier to a different data object}''~\cite{saml_core}.
Notice that this does not imply that identifiers must be generated via a cryptographic pseudorandom number generator (PRNG) --- only that the probability of collisions between identifiers generated by \emph{honest protocol parties} is negligibly low.
Clearly, this is a sensitive under-specified mechanisms since SAML identifiers are used to match responses to the original requests.

We examined several popular SAML frameworks (Shibboleth, opensaml, LemonLDAP, OneLogin) and  identified two cases (LemonLDAP and OneLogin PHP SDK) where a general purpose PRNG is used to generate SAML identifiers and, hence, the identifiers can be predicted by the adversary given enough past information.

We model two different behaviors corresponding to cryptographically strong and weak random identifiers: 
\begin{itemize}
    \item In the ``strongid'' model variants, the SAML identifiers are fresh nonces and, hence, unpredictable by the adversary.
    \item In the ``weakid'' model variants, the SAML identifiers are fresh nonces that are leaked to the adversary immediately upon generation (and before their use in the protocol). This corresponds the pedantic (but pessimistic) interpretation of the SAML standard --- no two honest parties will ever generate the same identifier, but, nevertheless, the adversary knows all the identifiers beforehand.
\end{itemize}

\subsection{Communication channels}
\label{section:tls}
SAML SSO profiles mandate the use of TLS channels while sending authentication request and responses.
Somewhat surprisingly (given the large number of case studies using the Tamarin prover), we found the previous approaches for modelling TLS communication in Tamarin lacking for our purpose and developed a simple novel approach for modelling the ``request-response over one-way authenticated TLS'' pattern in Tamarin.

Most common approach to modelling protected communication is by using \textit{secure} channels as in~\cite{tamarin_documentation}.
However, we found these to be too imprecise for our purpose.
Secure channels provide both authenticity and confidentiality, but they authenticate both parties (unlike one-way authenticated TLS used by SAML SSO).
Moreover, secure channels do not provide the link between requests and corresponding responses. 
A more suitable formalization is found in~\cite{tamarin_oidc_dunki_masters} --- the TLS sessions are created using a dedicated rule, messages are transferred through facts, and channel IDs are used to bind messages to specific TLS channels.
While this model of TLS communication is precise, we found it to be computationally unfeasible for our purpose.
The multiplicity of honest agent's interactions resulted in huge Tamarin precomputation times.

Our model of TLS communication is similar to one in~\cite{tamarin_oidc_dunki_masters}, except that we use ordinary network channels for communication and wrap TLS messages inside \emph{private function names}.
Using ordinary network channels significantly reduced precomputation time while wrapping TLS messages inside private functions makes it impossible for adversary to read the messages.
Furthermore, we do not require channel IDs as in~\cite{tamarin_oidc_dunki_masters} since TLS sessions in SAML SSO are short-lived (only one request and response).

TLS session is established between two \emph{agents} (the notion of agents is precisely described in the following Section~\ref{section:parties}).
For every TLS request towards an agent acting as server \msg{S}, a new session is created and labeled with unique \msg{tls\_session\_id}. 
Since we model one-way authenticated TLS, any agent can act as client and initiate such session.
However, only the agent \msg{S} can access the data in the request, and create a response bound to the same \msg{tls\_session\_id}. 
Such response can be accessed only by the agent that initiated the TLS session.
Note that the agent's name \msg{S} doubles as its TLS identity, in practice this can be, for example, its common name or a pinned certificate.

\begin{figure}[t]
\centering
\includegraphics[scale=0.55]{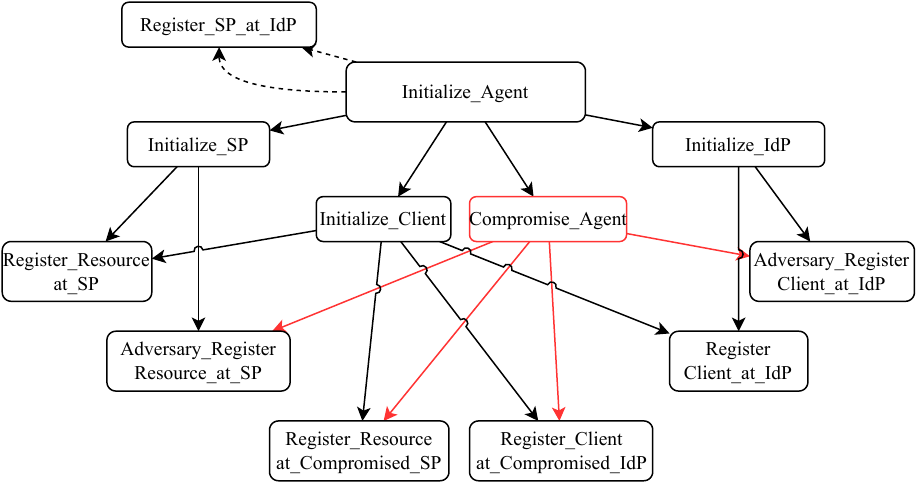}
\caption{
    Prerequisites Rules state machine.
    Initialized agent can become one of three honest parties or become compromised.
    Clients (and the adversary) can register resources at SPs (honest or compromised).
    Client (and the adversary) can register at IdPs (honest or compromised).
    Finally, any two SAML parties (SP and IdP) can establish trust.
    \label{figure:setup}
}
\end{figure}

\subsection{System setup}
\label{section:parties}
In our model, the system consists of a number of \emph{agents}.
There are three kinds of \emph{honest} agents -- the SPs and IdPs whose behavior is defined by the SAML specification (the \emph{SAML parties}) and the end user using a web browser client (the \emph{client}).
Additionally, adversary can create and control any number of \emph{compromised} agents.
In fact, since honest agents only communicate over TLS, the adversary must create and use compromised agents to interact with the system as described in Section~\ref{section:tls}.

Each agent in our model is assigned a unique \emph{name} --- a \emph{public value} in Tamarin.
For SAML parties, the agent name plays multiple roles --- we use as its domain name in the TLS protocol, as the URL at which the agent receives SAML requests, and as its SAML entity id (a globally unique name for a SAML party).
For clients, the agent name is purely a feature of the formal model, and each client is assigned a unique nonce \msg{clientID} during system setup.

\begin{figure*}[t]
\centering
\includegraphics[width=\textwidth]{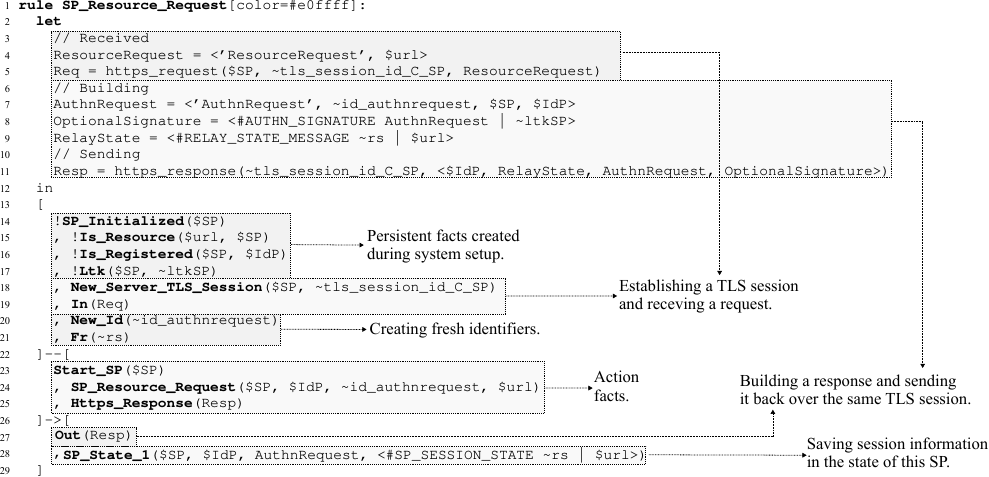}
\begin{lstlisting}[caption={Example Tamarin prover meta-rule --- SP recieves resource request --- annotated with information on the source and purpose of particular elements of the rule.},label={lst:saml_rule}]
\end{lstlisting}
\end{figure*}

\emph{System setup} rules model the agent initialization, key generation, role assignment, resource registration, and trust establishment. The rules are general and for allow arbitrary complex system and trust relationships.
A state machine illustrating the system setup rules is given in Fig.~\ref{figure:setup}.  
We assume unbounded number of  agents (SPs, IdPs, Clients) as well as unbounded number of parallel sessions for each agent.

\subsection{SAML parties -- the SP and the IdP}
Using the described methodology, building the formal model of the SP and the IdP to match the POST/Artifact use case is a straightforward but tedious process of translating SAML messages and processing rules into Tamarin rules.

As an example, in Listing~\ref{lst:saml_rule} we provide the rule modelling the behavior of SP when receiving a resource request from an honest agent or adversary (corresponding to the actions of the SP after receiving the very first request in Fig.~\ref{SAML_flow}). Technically, this is a \emph{meta-rule} as preprocessor directives (namely, \msg{\#AUTHN\_SIGNATURE}, \msg{\#RELAY\_STATE\_MESSAGE} and \msg{\#SP\_SESSION\_STATE}) are use to craft the exact rule for a given protocol variant. 

This is the initial step of protocol for SP, so this rule's pre-conditions contain facts modelling SP's initial state. These facts are \emph{persistent} --- they can be consumed many times, corresponding  to multiple sessions of the same SP.

\begin{itemize}
    \item \msg{!SP\_Initialized(\$SP)} --- persistent fact denoting that SP has been initialized.
    \item \msg{!Is\_Resource(\$url, \$SP)} --- persistent fact denoting that resource was registered at SP during system setup (possibly by the adversary).
    \item \msg{!Is\_Registered(\$SP, \$IdP)} --- persistent fact denoting that SP and a certain IdP have established trust during system setup.
    \item \msg{!Ltk(\$SP,\textasciitilde ltkSP)} --- persistent fact containing SPs private key created during SP initialization.
\end{itemize}
The Client has created a new TLS connection with unique TLS session ID in order to send request for resource located at \msg{\$url} at SP.
The TLS connection is denoted with the fact \msg{New\_Server\_TLS\_Session(\dots)} that SP consumes at this step.
The client's request $Req$ is delivered to this rule step through Tamarin built-in fact \msg{In()}. 
SP will consume the \msg{In(Req)} and create fresh values needed to build the authentication request that will be sent as response to client's request.
The creation of these fresh values is also a part of rule pre-conditions:
\begin{itemize}
    \item \msg{Fr(\textasciitilde rs)} --- fresh value denoting relay state is created in this rule step.
    \item \msg{New\_Id(\textasciitilde id\_AuthnRequest)} --- precondition modelling creation of a fresh value to use as the SAML identifier of the authentication request.
\end{itemize}
Finally, SP sends the authentication request along with the relaystate as a response $Res$ using Tamarin builtin fact \msg{Out()}, and creates the fact \msg{SP\_State\_1(\dots)} which denotes that SP is now in state where it received request for resource located at \msg{
\$url} from some client and that SP has sent an authentication request via the client to the chosen IdP.

As mentioned in Section~\ref{section:methodology}, \samlmsg{AuthnRequest}, \msg{RelayState}, and the rule creating the term containing the SAML ID --- $\msg{New\_Id(\textasciitilde id\_AuthnRequest)}$ --- vary according to the protocol variant.
For example, in the ``strongid'' variants the term is created by the rule that keeps the $\msg{\textasciitilde id\_AuthnRequest}$ secret, while in ``weakid'' variants it is sent to the network and, hence, made known to the adversary. 

\subsection{Client modeling}
\label{section:client_modeling}
In our work, the focus has been on thoroughly modelling the system parts defined by the SAML technical standards --- namely the SPs and IdPs.
Since behavior of the user and web client is not defined by SAML technical standards, we build the \emph{canonical} user and the web client.
We combine the end user and the web browser client into a single entity that we refer to as the \emph{client} in our formal model and in this paper.

The client is an idealized version of the end user and its web browser client that does not take into account all the intricacies of the browser behavior.
For example, our client will not follow arbitrary redirects, will not store cookies etc.
Also, in a more comprehensive model, one would need to take into account end user behavior and possible human errors.
For example, during authentication to the IdP we assume that the end user will authenticate and consent only when the SP matches the service provider the client initiated the session with.
Hence, we assume that the user is diligent enough to recognize the intended SP.

Moreover, the client will remember the SP it initialized the session with, and only again connect to that specific SP when it received the artifact from the SP.
In the real world, this is implemented using a combination of relay state and browser cookies.

    
In our formal model, the client can start the protocol flow with any service provider requesting any resource.
Hence, client can initiate sessions even with service providers it is not registered with and even with resources belonging to other clients.
This models possible mistakes or social engineering attacks on the end user.
    
Client authentication to the IdP is also not part of the SAML standard, so we created rules modelling simple client authentication.
Firstly, client registration at IdP during system setup (see Section~\ref{section:parties}) sets up a shared secret $clientSecret$ between the client and IdP.
While authenticating, after providing $clientSecret$ to the IdP over secure channel and after IdP successfully verifies it, the client is considered authenticated and IdP continues the protocol flow.

\subsection{Threat Model and The Adversary}
The threat model used in this paper closely matches the SAML threat model as identified by the SAML standard documents~\cite{saml_sec_privacy}. The only notable difference is that we are not considering denial-of-service attacks --- we feel that this is a complementary problem and not well suited for symbolic protocol analysis.

Tamarin prover comes equipped with the generic Dolev-Yao~\cite{dolev_yao} adversary that completely controls the network and is able to eavesdrop and, for example, launch man-in-the-middle or replay attacks. However, we model HTTPS requests and responses by private functions and, hence, the generic Tamarin adversary cannot construct them directly. Therefore, we introduce special rules that enable the adversary to send and receive HTTPS requests and responses, acting either as a \emph{server} and as a \emph{client}.

As in the SAML threat model, we assume that there could be a number of \emph{dishonest colluding} participants in the protocol. We model these by allowing the adversary to \emph{compromise} arbitrary agents. When an agent is compromised, the adversary gains access to its long term keys and secrets, and act as an HTTPS server using the compromised agent's identity.

Furthermore, the adversary can \emph{establish trust relationships} between compromised and other, honest, agents. Our model allows for arbitrary relationships. For example, an adversary can control a compromised client $C$ that has a resource  registered at an honest SP $S$, and user accounts at an honest IdP $I_1$ and at a compromised IdP $I_2$. Meanwhile, the SP $S$ could be configured to trust both IdPs $I_1$ and $I_2$. 
Initialization rules that enable the adversary to compromise protocol participants and setup trust relationships are denoted with red arrows in Fig.~\ref{figure:setup}.


\subsection{Limitations}
\label{section:limitations}

We make a number of simplifying assumptions to keep the complexity of the model manageable.
Most notable limitations are as follows.

\begin{description}
    \item[Client Fidelity]
    Clients are more idealized (compared to SPs and IdPs) and include both the end user and the web browser, as described in Section~\ref{section:client_modeling}. This simplified client model limits the scope of our analysis --- we will not discover attacks that arise from certain client-side behavior, such as browser-specific details (e.g., automatic redirects, cookie management), user errors
    (e.g., phishing or UI-based attacks), or the combination of these with the SAML protocol itself.
    Similarly, abstracting client authentication to IdP may impact our analysis by failing to discover attacks associated with the specific client authentication mechanisms.
    \item[Client Type Checking]
    Rules for client communication with SP and IdP are describing the behavior of a typical web browser.
    Aside from the initial request, the other two client rules are just automatic web browser responses to HTTP scripts that are embedded as SPs and IdPs requests and responses.
    This could have been implemented in formal model by just passing the message from the input to the output. In our model, however, we implemented type checking of messages forwarded by clients.  
    These rules are carefully designed, so the client is not comparing values of message variables, just checking variable type (fresh, public etc.).
    Using type checking, we removed a significant number of  message exchanges through client automatic rules and, therefore, reduced Tamarin reasoning time.
    Of course, such type checking is not used in real code implementations and our model is, therefore, less faithful in that regard.
    Nevertheless, we argue that type checks are not a significant loss of generality, since these messages are again type checked at SP and IdP side.
    Moreover, for cases where the adversary would like to send something that does not correspond to the message type, it can still do that through the adversary rules.
    
    \item[SAML SSO Profile 
    messages fields]
    SAML SSO Profile defines a number of properties and message definitions, message ids, name-spaces, additional assertion parameters etc., that we do not include in the formal model.
    These fields are well-specified in the protocol documentation and important in concrete implementations, but are not relevant to the security properties we consider in this paper.
    
    \item[SAML SSO Profile additional execution paths]
    Modeling too many execution paths can make the analysis intractable and the results difficult to interpret.
    Therefore, we focused on the critical aspects and security-relevant behaviors of the POST/Artifact use case and did not model all possible execution paths.
    Example execution paths that are not modeled are related to various error conditions (e.g., client sends an incorrect request).
    
    \item[Expressing security properties]
    In order to express all desired security properties (most notably the resource secrecy), we needed to add additional parameters not present SAML specification.
    While delivering resource to Client in the final step, SP generates a fresh unique variable which we called \emph{temporary secret} $ts$.
    Without this session unique value, we could not express property regarding the secrecy of resource delivered to the client in specific session.
    We did the same with assertion issued by IdP to SP where we inserted \emph{assertion secret} variable $sec$.  
\end{description}

\section{Security properties and analysis results}
\label{section:properties}
Now we present the security properties we consider in our analysis, their formalization as Tamarin lemmas, and the results of the automated verification.

\subsection{Formalization and verification of security properties}
\label{section:formalization}

Listing~\ref{lst:saml_lemma} presents two Tamarin lemmas --- one corresponding to the \emph{client resource authenticity} property and one corresponding to the strong variant of \emph{client resource secrecy} property --- we include them as representative examples to illustrate how security goals are expressed and verified within our formal model.
As mentioned in Section~\ref{section:tamarin}, we use exclusively trace properties, which are expressed as guarded first-order logic formulas. 
The lemma is true if the formula holds for all possible traces (executions) of the protocol. 

\begin{lstlisting}[caption={Example lemmas},label={lst:saml_lemma},float=t]
lemma sec_Client_Resource_Authenticity:
"All C S I cid csid url res ts #i.
  Client_Params(C, S, I, cid, csid, <res, url, ts>)@i
  ==>
  (Ex #j. Register_Resource_at_SP(C, S, cid, res, url)@j)
  | (Ex #j. Compromise_Agent(S)@j)
  | (Ex I1 #j #k. Compromise_Agent(I1)@j 
      & Register_SP_at_IdP(S, I1)@k)
  "

lemma sec_Client_Registered_Resource_Secrecy_Strong:
  "All C S url res id #i.
  Register_Resource_at_SP(C, S, id, res, url)@i
  ==>
  (not (Ex #j. K(res)@j))
  | (Ex #j. Compromise_Agent(S)@j)
  "
\end{lstlisting}

In both lemmas, the variables \msg{i}, \msg{j}, \msg{k} are typed as \emph{timestamps}, where other variables are untyped. Predicates \msg{Client\_Params}, \msg{Register\_Resource\_at\_SP}, etc. correspond to the action facts of the rules modeling SAML participants, while the \msg{Compromise\_Agent} corresponds to the setup rule that allows the adversary to compromise agents.
The predicate \msg{K} denotes the knowledge of the adversary.

Informally, the client resource authenticity lemma states the following: if a client \msg{C} (at some timestamp \msg{i}) completes a protocol session with the SP \msg{S} and the IdP \msg{I} with session parameters client ID \msg{cid}, obtaining resource \msg{res} at URL \msg{url} (line 3), then \msg{C} has (at some timestamp \msg{j}) registered that exact resource \msg{res} located at \msg{url} at the SP \msg{S} (line 5).
Furthermore, the lemma allows that \msg{C} has not registered the resource at the SP \msg{S} if one of the following holds: the SP \msg{S} is compromised (line 6), or there is some compromised IdP \msg{I1} that SP \msg{S} trusts (line 7).

The strong client registered resource secrecy lemma states the following: if a client \msg{C} registers a resource \msg{res} at the SP \msg{S} (line 13) then that resource remains unknown to the adversary --- it is never the case that the adversary knows the resource \msg{res} at some timestamp \msg{j} (line 15). Furthermore, the lemma allows that adversary learns \msg{res} if the specific SP \msg{S} is compromised (line 16).

Hence, in this lemma, we ask that secrecy holds even if there exists some compromised IdP \msg{I'} trusted by \msg{SP}. This lemma is included for sanity checking purposes as it should clearly fail --- a compromised IdP \msg{I'} trusted by \msg{S} can initiate a session, craft authentication responses for any client of its choice, and obtain their resources. 
Formal analysis validates this intuition and confirms that this lemma is falsified by an expected attack trace.

Note that the lemmas allow us to reason not only about the actions of participants in the protocol, and the knowledge of the adversary, but also express fine-grained honesty assumptions on the protocol participants by reasoning about events happened during the trust setup.

Tamarin proves trace properties by attempting to falsify them --- given a lemma, it tries to construct a trace that satisfies the negation of the property.
If such a counterexample is found, it represents a concrete attack. 
If no such trace exists and the search terminates, Tamarin produces a formal proof in the form of a proof tree, demonstrating that the property holds for all possible traces.
Since verifying trace properties is undecidable in general, Tamarin may not always terminate. 
In our analysis, we encountered several lemmas for which automated verification did not terminate or was prohibitively slow.
To address this, we restructured the model to reduce ambiguity, and introduced \emph{helper lemmas} to capture useful invariants, most notably related to TLS session and long term secrets. 
Additionally, we created proof-guiding oracles to direct the prover by prioritizing proof goals.
These were crafted based on the insights gained from attempts at manual verification of lemmas using the Tamarin's interactive mode. 

\subsection{Overview of security properties and analysis results}
\label{section:results_overview}
On a high level of abstraction, a Web SSO protocol should protect against unauthorized access and ensure confidentiality of sensitive user data. 
The exact desired security properties that the SAML Web SSO Profile aims to achieve are not explicit or precise in the official SAML documentation.
The ``Security considerations'' non normative document \cite{saml_sec_privacy} contains high-level descriptions of the threat model and the list of security techniques used to achieve specific low-level goals. 
For the SAML Web SSO profile, the standard~\cite{saml_sec_privacy} identifies several specific threats and  possible countermeasures, both within and outside the standard. 
For example, one identified threat is the ``Theft of the Bearer Token`` --- the adversary that obtains an artifact will be able to impersonate the user.
The countermeasures for this threat include using TLS to secure the connection, which is mandated by the standard. 
Finally, ~\cite{saml_sec_privacy} gives multiple practical guidelines for implementing systems based on the SAML framework in the secure manner.

Compared to SAML documentation, we consider a broader set of security properties and offer their precise formalization based on usual patterns used for formally modeling and verifying security protocols~\cite{Lowe_hierarchy, FKS_OAuth, FKS_OIDC, SAML_Armando_2013}. The properties considered in this paper encompass most of the threats described in~\cite{saml_sec_privacy} with denial-of-service attacks being the notable exception.

Most properties we consider can be grouped into three categories:
\begin{itemize}
\item \emph{Secrecy properties} state that the adversary cannot obtain information that is supposed to stay secret (e.g., the end-user resource or the details of the assertion).
In terms of~\cite{saml_sec_privacy}, these correspond to threats such as exposure of artifacts or assertions. Note that our secrecy properties are much more general than just the robustness against eavesdropping. For example, the adversary can attempt to compromise resource secrecy by launching man-in-the middle attacks, compromising various agents etc.
\item \emph{Agreement (or authentication)} properties state that the views of honest agents participating in the protocol session are compatible to some degree.
Agreement properties  differ when it comes to the details of \emph{what exactly} has to match in the compatible views of two protocol agents, and we consider all variants of authentication properties described in the seminal paper~\cite{Lowe_hierarchy} --- namely weak, non-injective and injective agreement. In terms of ~\cite{saml_sec_privacy}, these properties imply that the forged assertions, message insertion, message deletion, message modification, or man-in-the middle attacks cannot happen in the protocol.
\item{Freshness properties} 
state that certain actions (such as user authentication) must have happened recently in the past. 
In terms of ~\cite{saml_sec_privacy}, these properties cover the resistance to replay attacks.
\end{itemize}

Finally, as shown in table, properties can be grouped by the participant from whose point of view we consider the protocol execution (client, SP and IdP) and by the exact assumptions on entities compromised by the adversary.


In total, we have considered 28 security properties and have either formally verified or falsified every property for each of the 8 protocol variants.
The complete overview of the security properties and verification results is given in Table~\ref{table:results}.
The rest of this section gives an overview of the most relevant properties.

\begin{table*}[t]
\begin{center}
\setlength{\tabcolsep}{6pt}
\renewcommand{\arraystretch}{1.2}
\begin{tabular}{llcccccccc}
\multirow{2}{*}{\textbf{No.}} &
\multirow{2}{*}{\textbf{Property}} &
\multicolumn{8}{c}{\textbf{Protocol variant}} \\
& &  \textbf{srs} & \textbf{srw} & \textbf{sns} & \textbf{snw} & \textbf{urs} & \textbf{urw} & \textbf{uns} & \textbf{unw} \\
\hline
1 & sec\_Client\_Resource\_Authenticity & \yes & \yes & \yes & \yes & \yes & \yes & \yes & \yes \\
2 & sec\_Client\_Resource\_Authenticity\_Strong & \yes & \yes & \yes & \yes & \yes & \yes & \yes & \yes \\
3 & sec\_Client\_Registered\_Resource\_Secrecy & \yes & \yes & \yes & \yes & \yes & \yes & \yes & \yes \\
4 & sec\_Client\_Registered\_Resource\_Secrecy\_Strong & \no &   \no & \no & \no & \no & \no & \no & \no \\
5 & sec\_Client\_Resource\_Secrecy & \yes & \yes & \yes & \yes & \yes & \yes & \yes & \yes \\
6 & sec\_Client\_Resource\_Secrecy\_Strong & \yes & \yes & \yes & \yes & \yes & \yes & \yes & \yes \\
7 & sec\_Client\_Resource\_Freshness & \yes & \yes & \yes & \yes & \yes & \yes & \yes & \yes \\
8 & sec\_Client\_SP\_Non\_Injective\_Agreement & \yes & \yes & \yes & \yes & \yes & \yes & \yes & \yes \\
9 & sec\_Client\_SP\_Non\_Injective\_Agreement\_Strong & \yes & \yes & \yes & \yes & \yes & \yes & \yes & \yes \\
10 & sec\_Client\_SP\_Injective\_Agreement & \yes & \yes & \yes & \yes & \yes & \yes & \yes & \yes \\
11 & sec\_Client\_SP\_Injective\_Agreement\_Strong & \yes & \yes & \yes & \yes & \yes & \yes & \yes & \yes \\
12 & sec\_SP\_Client\_Resource\_Secrecy & \yes & \yes & \yes & \yes & \yes & \yes & \yes & \yes \\
13 & sec\_SP\_Client\_Resource\_Secrecy\_Strong & \yes & \yes & \yes & \yes & \yes & \yes & \yes & \yes \\
14 & sec\_SP\_Client\_Non\_Injective\_Agreement & \yes & \yes & \yes & \yes & \yes & \yes & \yes & \yes \\
15 & sec\_SP\_Client\_Non\_Injective\_Agreement\_Strong & \yes & \yes & \yes & \yes & \yes & \yes & \yes & \yes \\
16 & sec\_SP\_IdP\_Weak\_Agreement & \yes & \yes & \yes & \yes & \yes & \yes & \yes & \yes \\
17 & sec\_SP\_IdP\_Weak\_Agreement\_Strong & \yes & \yes & \yes & \yes & \yes & \yes & \yes & \yes \\
18 & sec\_SP\_IdP\_Non\_Injective\_Agreement & \yes & \yes & \yes & \yes & \yes & \yes & \yes & \yes \\
19 & sec\_SP\_IdP\_Non\_Injective\_Agreement\_Strong & \yes & \yes & \yes & \yes & \yes & \yes & \yes & \yes \\
20 & sec\_SP\_IdP\_Authentication\_Freshness & \yes & \yes & \yes & \yes & \yes & \no & \yes & \no \\
21 & sec\_SP\_IdP\_Authentication\_Freshness\_Strong & \yes & \yes & \yes & \yes & \yes & \no & \yes & \no \\
22 & sec\_SP\_IdP\_Assertion\_Secrecy & \yes & \yes & \yes & \yes & \yes & \yes & \yes & \yes \\
23 & sec\_SP\_IdP\_Assertion\_Secrecy\_Strong & \yes & \yes & \yes & \yes & \yes & \yes & \yes & \yes \\
24 & sec\_IdP\_SP\_Non\_Injective\_Pre\_Agreement & \yes & \yes & \yes & \yes & \no & \no & \no & \no \\
25 & sec\_IdP\_SP\_Weak\_Agreement & \yes & \yes & \yes & \yes & \yes & \yes & \yes & \yes \\
26 & sec\_IdP\_SP\_Non\_Injective\_Agreement & \no & \no & \no & \no & \no & \no & \no & \no \\
27 & sec\_IdP\_SP\_Injective\_Agreement & \no & \no & \no & \no & \no & \no & \no & \no \\
28 & sec\_IdP\_SP\_Assertion\_Secrecy & \yes & \yes & \yes & \yes & \yes & \yes & \yes & \yes \\
\hline 
\multicolumn{10}{l}{\textbf{Legend:}} \\
\multicolumn{10}{l}{\quad first letter of the protocol variant -- 
authorization requests are signed (``s'') or unsigned (``u'')} \\
\multicolumn{10}{l}{\quad second letter of the protocol variant -- relay state mechanism is used (``r'') or not used (``n'')} \\
\multicolumn{10}{l}{\quad third letter of the protocol variant -- random IDs are cryptographically strong (``s'') or weak (``w'')} \\
\multicolumn{10}{l}{\quad \yes -- property verified, \no -- property 
 falsified} \\
\end{tabular} 
\caption{Analysis results for each property and protocol variant. }
\label{table:results}
\end{center}
\end{table*}


\subsection{Client PoV}
\label{section:clientpov}
First, we investigate the security properties from the client's point of view.
That is, we assume that a client that has successfully completed a protocol session apparently with the SP $S$ and the IdP $I$, and investigate what guarantees hold for the client under different assumptions on honesty of other protocol principals.
The honesty assumptions we consider are as follows:
\begin{itemize}
    \item Both the service provider $S$ and the identity provider $I$ used in this particular session are honest. 
    \item The service provider $S$ used in this particular session is honest, and additionally the service provider $S$ has only established trust with honest identity providers. 
\end{itemize}
Hence, whenever it made sense, we considered two versions of security lemmas: the \emph{strong} and the \emph{weak} version corresponding to the first and second assumption above.
By considering the two versions, we are aiming to measure the damage to the protocols in scenarios where a service provider establishes trust with an identity provider which happens to be compromised. 

\begin{description}
    \item[Client resource authenticity]
    If a client $C$ completes a protocol session with the SP $S$ and the IdP $I$ with session parameters client ID $cid$, obtaining resource $res$ at URL $url$ then $C$ has registered that exact resource $res$ located at $url$ at the SP $S$. 
\end{description}

As mentioned above, we also consider two versions of property --- the \emph{weak} and the \emph{strong} version.
In the stronger version, instead of assuming that $S$ does not trust any compromised identity provider $I'$, we only assume that the specific identity provider $I$ used in this particular session is not compromised.
Let us elaborate a bit more: assume a deployment where the service provider $S$ has established trust with three different identity providers $I_1$ , $I_2$ and $I_3$ such that $I_3$ is compromised while the other two identity providers are honest. 
If the strong version of \emph{client resource authenticity} holds, that means that the resource provided in a particular session is authentic as long as an honest IdP (e.g., either $I_1$ or $I_2$ was used \emph{in that particular session}).
If only the weaker variant of the property holds (and stronger does not), that means that the \emph{client resource authenticity} cannot be guaranteed in this deployment -- even if an honest IdP is used in a particular session, the adversary could still violate the resource authenticity property by launching an attack (that somehow takes advantage of the fact that IdP $I_3$ is dishonest).

The formal analysis shows that this security property holds for both weak and strong version for all tested protocol variants. Hence, in the SP-initiated SSO with POST/Artifact use case, the client is guaranteed that the resource he is provided with is the resource that it registered as long as the SP and IdP used in the session are both honest.

\begin{description}
    \item[Client registered resource secrecy]
    If a client $C$ registers a resource $res$ at the SP $S$ then that resource remains forever unknown to the adversary.
\end{description}

As described in~\ref{section:formalization} the strong variant is included for sanity checking purposes, and it fails for all protocol variants, as expected.

The formal analysis shows that the weaker version of the property holds for all protocols considered

Note that, this property considers the secrecy of the long-lived resources across sessions.
The following property considers the resource secrecy on the session level.
\begin{description}
    \item[Client resource secrecy]
    If a client $C$ completes a protocol session with the SP $S$ and IdP $I$ with parameters client ID $cid$, and resource $res$ located at URL $url$ accompanied by temporary secret $ts$, then the temporary secret $ts$ remains forever unknown to the adversary.
\end{description}

Temporary secret $ts$ is of type fresh and is created and delivered in the last step when SP delivers resource to client.
This property means that if the client manages to complete the protocol and to acquire a resource in that concrete protocol run, it is not possible that the adversary obtained the resource compromising that particular session.

The formal analysis shows that this security property holds for both weak and strong version for all protocol variants.

\begin{description}
    \item[Client resource freshness]
    If a client $C$ with ID $cid$ completes the protocol session with the SP $S$ obtaining resource $res$ accompanied by a temporary secret $ts$, then the temporary secret $ts$ was created fresh after $C$ has requested the resource.
    This property holds unless $S$ is compromised. 
\end{description}
Informally this property states that the client cannot be tricked into accepting an old resource.
This property does not depend on the identity provider and, hence, we do not make any assumptions on the honesty of IdPs present in the system.

The formal analysis shows that this security property holds for all protocol variants.

\begin{description}
    \item[Client -- SP non-injective agreement]
    If a client $C$ with ID $cid$ completes the protocol session with the SP $S$ and IdP $I$ obtaining resource $res$ at $url$ with temporary secret $ts$, then $S$ has completed a protocol with $cid$ and the session parameters (the resource $res$, URL $url$ and temporary secret $ts$) all match. 
\end{description}

This is a standard two party authentication property --- if a client $C$ has completed a session with $S$ then there is a session of $S$ where the session parameters agree with those in the client session. 

The formal analysis shows that this security property holds for both weak and strong version for all protocol variants.

The following property requires that, additionally, there are no multiple sessions with the same session parameters.
\begin{description}
    \item[Client -- SP injective agreement]
    If a client $C$ with ID $cid$ completes the protocol session with the SP $S$ and IdP $I$ obtaining resource $res$ at URL $url$ with temporary secret $ts$ then $S$ has completed a protocol with client represented by $cid$ and the session parameters match and, additionally, there must not exist a client session of with same parameters $res$ and $cid$.
\end{description}

The formal analysis shows that this security property holds for both weak and strong version for all protocol variants.

\subsection{Service Provider PoV}
\label{section:sppov}
From the SP's PoV, we considered similar security properties as for client's PoV. In addition to authentication properties between the client and SP, we consider authentication properties between SP and IdP. Furthermore, we also consider resource and assertion secrecy.
As before, wherever applicable, we consider the weak and the strong version of the property depending on the exact assumption on the honest IdPs.

The analysis shows that the resource secrecy (both weak and strong) and the non-injective agreement (again both weak and strong) between the SP and the client hold for all protocol variants.

It is important to note that these properties between the client and the SP are not redundant or implied by the analogous properties from the PoV of the client. 
For example, \emph{SP -- client resource secrecy} property says that the temporary secret $ts$ served by the SP $S$ in some session remains unknown to the adversary as long as the client that registered the corresponding resource and the IdP $I'$ used in that particular session are not compromised.
This property guarantees that the honest client's resources remain secret even in the scenarios when the client mistakenly establishes or completes sessions with compromised service providers --- and these scenarios are not covered by the analogous properties from the PoV of the client.  
We do not include the \emph{SP -- client injective agreement} in the list of properties since it is trivially implied by the non-injective agreement --- the session parameters (e.g., the ID $ida$ and the temporary secret $ts$ are generated fresh by $S$ making the SP's session trivially unique).

Taken together with verified \emph{client -- SP injective agreement} positive results, we prove that SP-initiated SSO with POST/Artifact binding offers strong bidirectional agreement between the client and the SP on the session level as long as the IdP used in the session is honest.

Next, we consider authentication properties between the SP and the IdP (namely the weak agreement and the non-injective agreement), and show that the weak and strong versions of both properties hold for all protocol variants.
The agreement between the SP and the IdP is probably the most important property of the protocol as it captures the purpose of the SSO --- if an SP accepts the assertion from an IdP, then IdP must have issued the assertion after authenticating the client.
In all properties between SP and IdP we make no assumptions on the honesty of the clients. Hence, the properties hold when all clients are under the control of the adversary.

In the following property descriptions, $ida$ is the unique identifier of an \samlmsg{AuthnRequest} message while $mh$ is an unique message handle of an \samlmsg{Artifact} and $idr$ is unique identifier of an \samlmsg{ArtifactResolve} message.
\begin{description}
    \item[SP -- IdP non-injective agreement]
    If an SP $S$ completes a protocol with the client with ID $cid$ and the IdP $I$, then $I$ had completed a protocol and authenticated the client with ID $cid$. Furthermore, and all other parameters ($mh$, $ida$ and $idr$) match.
\end{description}
Again, injective agreement property is trivially implied by the non-injective  agreement is, hence, not included in the analysis. We do, however, consider one additional agreement property that, informally, requires that the IdP has authenticated the client \emph{recently}.

\begin{description}
    \item[SP -- IdP authentication freshness]
    In addition to the non-injective agreement above, this property requires that the authentication of the client with ID $cid$ is fresh, in the sense that it happened after the $S$ has sent the authentication request.
\end{description}

The analysis verifies these properties for 6 protocol variants and falsifies the property for 2 protocol variants.
The property does not hold in variants where the authentication requests are not signed, and the IDs are cryptographically weak.
In the attack that violates this property, the adversary guesses the ID used for the authentication request and uses it to construct an  authentication request that appears to be originating from $S$. 
The result is that the client authentication is triggered before the genuine authentication request is issued by the SP $S$.
We do not consider the violation of this property to be a serious drawback of the protocol in practice, as discussed in Section~\ref{section:discussion}.

Finally, we consider the secrecy of assertions accepted by the SP. 
Authentication assertion can possibly hold some sensitive user data, and therefore its secrecy is also important in some deployments.

\begin{description}
\item[SP -- IdP assertion secrecy]
    If an SP $S$ accepts an assertion containing an assertion secret $sec$, then that assertion remains unknown to the adversary.
\end{description}
Formal analysis shows that both the strong and the weak version of the property holds in all protocol variants.

\subsection{Identity provider POV}
\label{section:idppov}
From the IdP's PoV, we consider several authentication properties between IdP and SP as well as the secrecy of assertions issued by the IdP. In all properties we assume that the SP $S$ is honest and, again, make no assumptions on the honesty of clients in the protocol.

\begin{description}
    \item[IdP -- SP non-injective pre-agreement]
    If an IdP $I$ has completed the protocol and issued an assertion to the SP $S$ in response to an \samlmsg{AuthnRequest} with ID $ida$, then $S$ has issued the \samlmsg{AuthnRequest} with ID $ida$ and has chosen $I$ as the IdP for the session.
\end{description}

Formal analysis verifies these properties for the four protocol variants where the authentication request are required to be signed and falsifies the property for the four protocol variants where the authentication requests are unsigned.

This is not too surprising --- if an IdP accepts unsigned authentication requests, then the adversary can issue them arbitrarily.
However, the exact failure mode is more subtle since it requires that $I$ completes the protocol, which involves resolving the artifact, which, in turn, requires mutual authentication between the SP $S$ and the IdP $I$.

In the attack falsifying the property, the adversary acting as a client obtains a genuine \samlmsg{AuthnRequest} from the SP $S$, but replaces the identifier $ida$ with a different value before forwarding the request to the $I$.
The $I$ will, now, complete the protocol --- it will resolve the artifact even though there is a mismatch between the $ida$ parameters in the sessions of $I$ and $S$.

Note that the SP $S$ will not complete the protocol session --- in the last step it will reject the assertion since the \samlmsg{InResponseTo} field in the assertion does not match the original \samlmsg{AuthnRequest} ID $ida$. 
This behavior agrees with the assertion processing rules in the SAML specifications.
Again, we do not consider this failure to have a significant impact in real world deployments.

\begin{description}
    \item[IdP -- SP weak agreement]
    If an IdP $I$ has completed the protocol and issued an assertion to the SP $S$, then $S$ has issued the corresponding \samlmsg{AuthnRequest} choosing some $I'$ as the identity provider, and also requested the \samlmsg{Artifact} that $I$ resolved.
\end{description}
It is straightforward to conclude that weak agreement should hold due to mutual authentication during the artifact resolution phase.
Formal analysis indeed shows that this property holds in all protocol variants.

\begin{description}
    \item[IdP -- SP non-injective agreement]
    If an IdP $I$ has completed the protocol and issued an assertion to $S$, then $S$ has issued the corresponding \samlmsg{AuthnRequest} choosing $I$ as the identity provider and also requested the \samlmsg{Artifact} that $I$ resolved. Hence, $I$ and $S$ agree on the parameters of the \samlmsg{AuthnRequest} and the \samlmsg{Artifact}.
\end{description}
Formal analysis falsifies this property for every protocol variant.
This is somewhat surprising at the first glance, but deeper analysis reveals that the property should indeed fail.

In the protocol execution falsifying this property, the adversary takes the role of a compromised client and performs what we call an \emph{artifact-session confusion attack} as follows:
\begin{enumerate}
    \item Adversary starts a first session with the SP $S$ requesting the resource $res_1$ and obtaining a genuine \samlmsg{AuthnRequest} with ID $ida_1$.
    \item Adversary starts a second session with the SP $S$ requesting the resource $res_2$ and obtaining a genuine \samlmsg{AuthnRequest} with ID $ida_2$.
    \item Adversary continues the second session, authenticates with IdP $I$ and obtains a genuine \samlmsg{Artifact} with handle $mh_2$.
    \item Adversary continues the first session and sends the \samlmsg{Artifact} with handle $mh_2$ to $S$.
    \item SP $S$ resolves the received \samlmsg{Artifact}, causing the $I$ to complete the first session resolving \samlmsg{Artifact} with handle $mh_2$ for \samlmsg{AuthnRequest} with ID $ida_1$.
\end{enumerate}
The end result of the execution is that IdP has completed a protocol session, but there is no session of $S$ that agrees on the parameters of \samlmsg{AuthnRequest} and \samlmsg{Artifact}.
This vulnerability allows an adversary participating in a second session to manipulate the Service Provider (SP) into resolving an artifact from the first session.
It is important to clarify that this vulnerability does not elevate the adversary's privileges to completely bypass authentication --- the adversary must still authenticate with the Identity Provider (IdP) to obtain the artifact in the second session. 


This attack is consistent with SAML specification for the following reasons: First, there is no way for $S$ to check that the \samlmsg{Artifact} received in step 4 corresponds to a different session -- the artifact is opaque by design and the only contains the information about its originator $I$. Second, it is neither required or possible for $I$ to perform additional checks during the artifact resolution protocol in step 5.  We were also able to confirm this attack is possible in practice by launching it against our simple deployment consisting of Shibboleth~\cite{shibboleth_doc} Service Provider v3 and Identity Provider v4.  

Again, the damage to the protocol is limited due to the fact that  a SAML compliant SP $S$ \emph{will not} complete the protocol and issue the resource --- it will abort the protocol after detecting the mismatch between the received assertion and the original \samlmsg{AuthnRequest}. 
Specifically, the standard states that the SP must verify that the \samlmsg{InResponseTo} attribute in the bearer \samlmsg{SubjectConfirmationData} equals the ID of its original \samlmsg{AuthnRequest} message, unless the response is unsolicited ~\cite{saml_profiles} --- and our model implements this.
Hence, we believe that this attack has \emph{no direct impact} on the security of deployments, but it highlights the necessity of following the assertion processing rules specified in the SAML standard.

Finally, we consider the secrecy of assertions from the PoV of the IdP.
\begin{description}        
    \item[IdP -- SP assertion secrecy]
    The \samlmsg{Assertion} that IdP $I$ issued to the service provider $S$ remains unknown to the adversary.
\end{description}
Formal analysis verifies this property for every protocol variant. Taken together with the analogous property from the PoV of the SP, we prove that SP-initiated SSO with POST/Artifact binding using case guarantees assertion secrecy even if the assertions are not encrypted.

\subsection{Challenges of automated verification}
\label{section:verification}

\begin{table*}[t]
\begin{center}
\setlength{\tabcolsep}{6pt}
\renewcommand{\arraystretch}{1.2}
\begin{tabular}{llrrrrrrrr}
\multirow{2}{*}{\textbf{No.}} &
\multirow{2}{*}{\textbf{Lemmas}} &
\multicolumn{8}{c}{\textbf{Protocol variant}} \\
& &  \textbf{srs} & \textbf{srw} & \textbf{sns} & \textbf{snw} & \textbf{urs} & \textbf{urw} & \textbf{uns} & \textbf{unw} \\
\hline
1 & Executable lemmas & 121s & 121s & 120s & 104s & 126s & 116s & 101s & 98s \\
2 & Helper lemmas & 104s & 104s & 98s & 97s & 103s & 103s & 96s & 96s \\
3 & Client PoV properties & 5813s & 5810s & 3042s & 3035s & 6382s & 6396s & 2976s & 2981s \\
4 & SP and IdP PoV properties & 425s & 431s & 403s & 403s & 1569s & 242s & 253s & 230s \\
\hline 
\end{tabular} 
\caption{Total analysis times (in seconds) for all lemma groups and protocol variants.}
\label{table:times}
\end{center}
\end{table*}

The Tamarin meta model (with comments removed) contains 812 lines of code, about half of which correspond to protocol rules and half to lemmas and security properties. 
Each of the 8 concrete models for protocol variants is of roughly the same size as the meta model.
Even though the models are of moderate size, the analysis proved challenging with many proofs never terminating, even in the simplified setting with a bounded number of sessions.
Hence, several insights were needed to achieve termination and to finally verify/falsify all properties in a general setting with an unbounded number of agents and sessions.

Final analysis was performed using Tamarin version 1.6.1 on an 8-core/16-thread desktop machine with Intel i9-9900K CPU and 32G of RAM. The total analysis time for all properties and protocol variants is close to 10 hours (160 hours of CPU time); the breakdown per protocol variant and the type of lemma is given in Table~\ref{table:times}.

The subsequent subsections elaborate on the key lessons learned that are essential for achieving finite-state verification.

\subsubsection{Combinatorial Explosion}
The Tamarin prover exhaustively explores all possible rule applications within a protocol to reach the desired pre-conditions.

One source of complexity were the TLS channel rules initially implemented through facts --- similar how TLS channels have been modelled in Tamarin before~\cite{tamarin_oidc_dunki_masters, tamarin_oidc_xenia_masters} (which is, in turn, similar to confidential and authentic channels in the Tamarin documentation).
This approach, however, resulted in exploding complexity during the precomputation phase due to the enumeration of all possible interactions of honest protocol participants and the adversary via the TLS channels.
We addressed this complexity by, first, modelling TLS channels using a combination of private function symbols and adversary rules as described in Section~\ref{section:tls}. This, along with a \emph{source} lemma to resolve partial deconstructions, significantly reduced the complexity of precomputation (while, of course, moving some of the complexity to the main analysis phase). Secondly, we proved a number of TLS-related helper lemmas, further reducing the complexity of the analysis.


Another source of complexity is the large number of fresh values used in the protocol ---most SAML messages require IDs which are modelled as fresh values.
In each of our protocol models, there were roughly 50 refined sources for the $KU(\sim{}t)$ fact which translates to 50 new branches in the search tree each time a goal of such type is processed.
We finally tamed the complexity by writing proof-guiding oracles for each problematic lemma, or groups of related lemmas.

\subsubsection{Protocol and property abstraction}
To achieve finite-state verification, security properties and/or protocol rules can be strategically and carefully abstracted. This reduces the complexity of the verification process, enabling it to complete within a reasonable time-frame.

Also, we significantly improved the analysis time of the security properties considered from the SP's and the IdP's PoV by removing all client rules from the model.
Since there is no assumptions on the honesty of clients in those properties and, furthermore, all client actions can be performed by the adversary (as evidenced by the executable lemmas), there is no loss of generality here.

Finally, when running sanity checking tests (the executable lemmas) we restricted the model to a bounded number of sessions (only one client, one SP and one IdP) and achieved much better analysis times.

\section{Summary and discussion of main results}
\label{section:discussion}
Based on the results of the analysis, we make the following claims on the POST/Artifact use case:

\subsection{The POST/Artifact use case satisfies desired security goals for a Web SSO system.}
Formal analysis verifies the most important security properties for all the considered protocol variants. 
Some security properties were falsified for some protocol variants, however, these failures are either expected or do not pose a significant threat in the real-world deployments of the use case.

The violation of the \emph{SP-IdP Authentication Freshness} property in some protocol variants, means that the IdP can authenticate clients before the SP requests their authentication. 
However, SAML assertions contain additional attributes that can, and should, be used when the time of authentication is of importance. For example, the \samlmsg{AuthnInstant} attribute specifying the time at which the authentication took place or the \samlmsg{NotBefore}/\samlmsg{NotOnOrAfter} attributes specifying the validity time period of the assertion.

Additionally, the use case provides a certain level of robustness against a compromised IdP.
Of course, a compromised IdP may still impersonate and access resources of any user at any trusting SP (as we demonstrate by falsifying the strong version of the \emph{Client registered resource secrecy} property).
However, a compromised IdP (even if trusted by SPs) cannot degrade security of protocol sessions where an honest IdP is used.

\subsection{Details left unspecified by the SAML standard can have an impact on the exact security properties provided by the use case.}
The protocols with cryptographically strong IDs have slightly different security properties than those with weak IDs.
Admittedly, the difference is subtle, only present when combined with unsigned authentication requests, and unlikely to have any practical consequences.
However, we believe the results validate the general concern that the seemingly purely functional choices left to developers by the SAML standard can end up having an impact to the security of the implemented system. However, the formal analysis did not discover any difference between the two different relay state mechanisms considered.

\subsection{There is no clear mapping between SAML optional protection mechanisms and provided security guarantees.}
Adding signatures to authentication requests does not have a significant influence on the security of the use case.
For the most important security properties, there is no difference between the variants where signatures are required and when they are not.
Besides the above-mentioned authentication freshness issue, the only difference is the \emph{IdP-SP non-injective pre-agreement} property, where the adversary can make the IdP issue unsolicited authentication responses (that the SP will ultimately reject). This claim should be taken only in the limited context of symbolic protocol analysis --- in a real-world deployment an IdP that requires signed authentication requests may be much less vulnerable to various low-level attacks (compared to an IdP that does not) simply due to a reduced attack surface as described in~\cite{saml_sec_privacy}.

The fact that unsigned authorization requests do not result in serious drawbacks can be explained by the fact that two-way authentication is established during artifact resolution before the assertion is delivered.

Additionally, this use case provides assertion secrecy even though no assertion encryption mechanism is used.
Again, this fact can be explained by the artifact resolution mechanism.

Hence, it is hard for SAML deployers to decide what optional security mechanisms to include, since it is highly non-trivial to deduce the impact of those security mechanisms on the final SSO system. 

\subsection{Artifact-session confusion attack mitigation}
The artifact-session confusion attack breaking the \emph{IdP-SP non-injective agreement} property is interesting, but the extent of the attack is that an IdP can be made to resolve authentication responses that \textit{standards compliant} SP will never consume.
Hence, we believe that there is no practical impact of this attack, but repeat that care must be taken by the SP to follow the assertion processing rules completely.
However, we do propose a  modification to the artifact creation process defined within the SAML standard \cite{saml_core} that will mitigate the attack.

The existing standard mandates that artifacts be resolvable to their originating sender and exhibit single-use semantics.
While SAML standard says that ``Different types can be defined and used without affecting the binding'' (Section 3.6.4 in \cite{saml_bindings}), it currently defines a single artifact format consisting of a random message handle and the \samlmsg{SourceId} uniquely identifying the ``artifact issuer identity and the set of possible resolution endpoints``. The artifact receiver uses the \samlmsg{SourceId} to decide where to resolve the artifact. Note that \samlmsg{SourceId} processing plays a significant role for the security of the use-case --- we discovered a man-in-the-middle attacks in the early version of the formal model where the \samlmsg{SourceId} was missing from the artifact.

Our proposed mitigation for the artifact-session confusion attack is to extend the artifact format and include a reference to the original SAML request within the artifact.
Specifically for this use case, the artifact would contain the `id\_authnrequest` identifier originating from the SAML Authentication Request sent by the SP to the IdP.

This approach strengthens session integrity by enabling the SP and IdP to verify the consistency between the received artifact and the corresponding SAML request.
Even if the IdP receives an artifact containing a mismatch between the message handle and the `id\_authnrequest` identifier, the \samlmsg{ArtifactResolve} message can be discarded, effectively thwarting artifact-session confusion attacks.

We formally model this proposed fix and formally verify that it does indeed solve the problem. We skip the details in the paper, but in~\cite{github_code_ref} we make available the protocol model with the proposed fix and its complete analysis. 
These results demonstrate that "IdP – SP non-injective pre-agreement," "IdP – SP non-injective agreement," and "IdP – SP injective agreement" properties all hold true for in all eight variants with the proposed fix.

\section{Conclusions and future Work}
\label{section:conclusion}
We take a step towards comprehensive formal analysis of SAML V2.0 Web Browser SSO Profile by proposing a simple methodology and use it to formally verify an \emph{SP-initiated SSO with POST/Artifact Bindings} use case.

Before building the model, we identify sensitive optional features and sensitive under-specified mechanisms and analyze eight different protocol variants to test if seemingly functional choices can impact the security of the proposed protocol.
The model allows for almost trust relationships, and includes a Dolev-Yao adversary that completely controls the network and can compromise arbitrary participants. 

We consider a comprehensive set of authentication, secrecy and freshness properties and, through machine-verifiable formal analysis, show that the considered use case does indeed provide the most important security properties expected of a Web SSO protocol. 

We do discover minor weaknesses, including the artifact-session confusion attack which breaks the \emph{IdP-SP non-injective agreement} property.
We argue that this vulnerability holds no practical impact when the SP implements SAML in accordance with standard.
Nevertheless, we suggest a slight modification to the standard and formally verify that it does indeed address the discovered weakness.

In future work we would like to create comprehensive formal models of all possible use cases of \emph{Web Browser SAML SSO Profile} and to automatically verify presented security properties.
In such work we would like to include analysis of concrete SAML SSO implementations and to compare those implementations with our formal model.

\bibliographystyle{IEEEtran}

\begin{IEEEbiography}[{\includegraphics[width=1in,height=1.25in,clip,keepaspectratio]{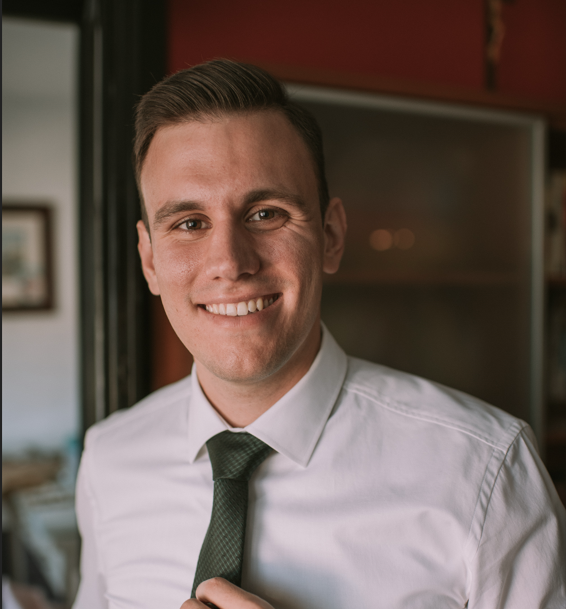}}]{Zvonimir Hartl} received the Master of Science in Information and Communication Technology degree from the University of Zagreb, Faculty of Electrical Engineering and Computing, Croatia, in 2019, where he is currently pursuing the Ph.D. degree. He currently works at Infobip, a Croatian company specializing in omnichannel communication. His research interests include formal modeling of security protocols.
\end{IEEEbiography}

\begin{IEEEbiography}[{\includegraphics[width=1in,height=1.25in,clip,keepaspectratio]{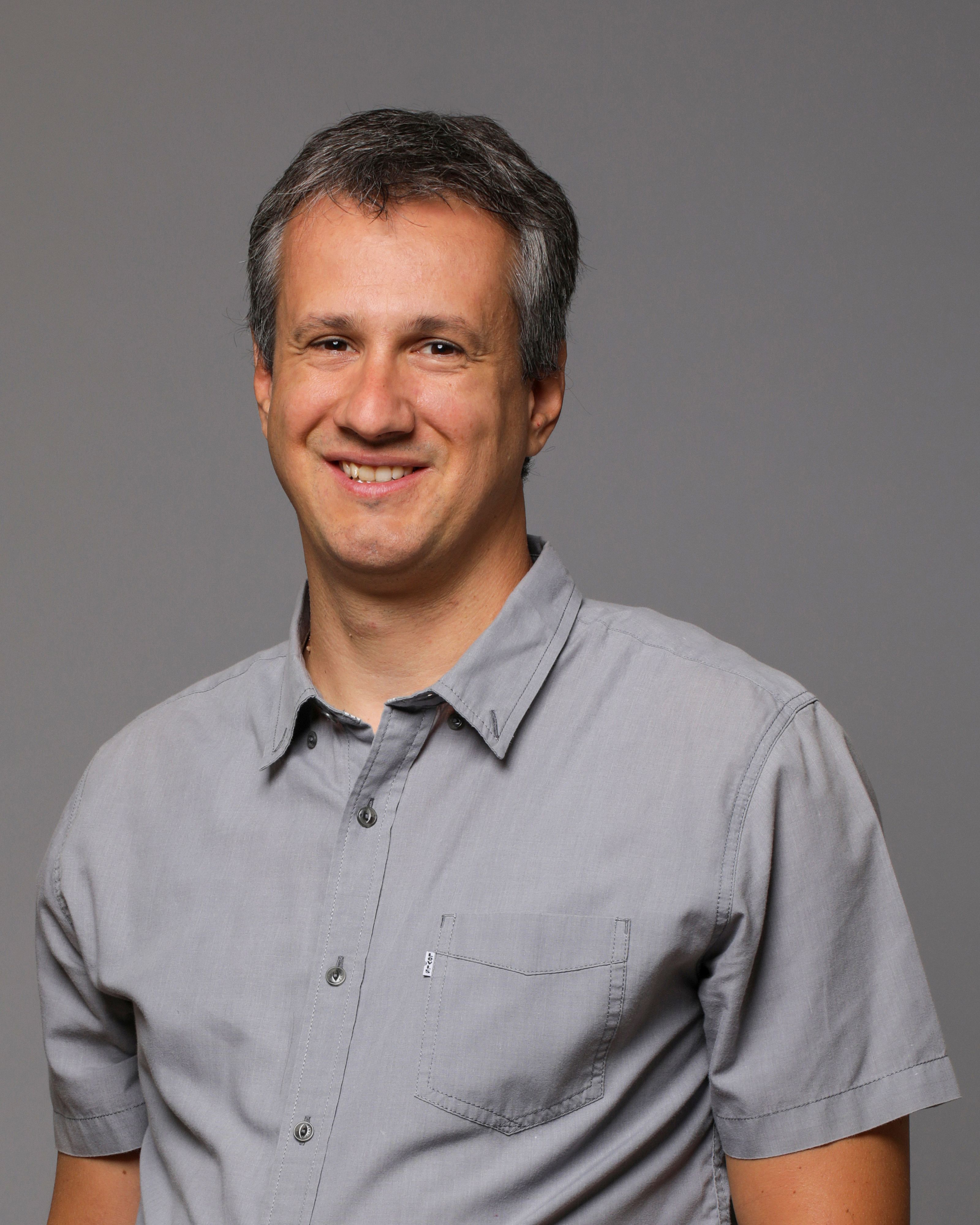}}]{Ante Derek} is an Associate Professor at the Faculty of Electrical Engineering and Computing, University of Zagreb. He participates in a number of national and EU-funded projects in the area of computer security. His research interests are in the area of applying formal methods to problems in computer security, privacy, and cryptography.
\end{IEEEbiography}

\EOD

\end{document}